# The REESSE2+ Public-key Encryption Scheme

Shenghui Su and Shuwang Lü

**Abstract**: This paper gives the definitions of an anomalous super -increasing sequence and an anomalous subset sum separately, proves the two properties of an anomalous super-increasing sequence, and proposes the REESSE2+ public-key encryption scheme which includes the three algorithms for key generation, encryption and decryption. The paper discusses the necessity and sufficiency of the lever function for preventing the Shamir extremum attack, analyzes the security of REESSE2+ against extracting a private key from a public key through the exhaustive search, recovering a plaintext from a ciphertext plus a knapsack of high density through the $L^3$ lattice basis reduction method, and heuristically obtaining a plaintext through the meet-in-the-middle attack or the adaptive-chosen-ciphertext attack. The authors evaluate the time complexity of REESSE2+ encryption and decryption algorithms, compare REESSE2+ with ECC and NTRU, and find that the encryption speed of REESSE2+ is ten thousand times faster than ECC and NTRU bearing the equivalent security, and the decryption speed of REESSE2+ is roughly equivalent to ECC and NTRU respectively.

## I. Introduction

Merkle and Hellman proposed a public-key cryptosystem based on the knapsack problem [1] in 1978, the second year after Diffie and Hellman had delivered their pioneering paper "New Directions in Cryptography" [2].

Let $\{a_1, \ldots, a_n\}$ be a positive integer sequence, and if $a_i$ satisfies $a_i > \sum_{j=1}^{i-1} a_j$ for $i = 2, \ldots, n$, $\{a_1, \ldots, a_n\}$ is called a super-increasing sequence.

Any positive integer sequence $\{c_1, \ldots, c_n\}$ may be called a knapsack. Especially, in the MH scheme, $\{c_1, \ldots, c_n \mid c_i \equiv a_i w \ (\% \ M)\}$ is a public key, where $M > \sum_{i=1}^{n} a_i$ is a modulus, and $W < M$ is a transform parameter. Assume that the integers $x_1, \ldots, x_n \in [0, 1]$ is a plaintext, the subset sum $y \equiv \sum_{i=1}^{n} c_i x_i \ (\% \ M)$ is a ciphertext, and then how to solve $\{x_1, \ldots, x_n\}$ to $y \equiv \sum_{i=1}^{n} c_i x_i \ (\% \ M)$ is called the knapsack problem.

The super-increasing sequence $\{a_1, \ldots, a_n\}$ has a weakness that $a_{i+1}$ is roughly double $a_i$. Shamir, one of the inventors of the RSA cryptosystem [3], grasped this weakness and in polynomial time extracted a related private key from a public key through the convergence of infinitesimal points in 1982 [4]. Extracting the private key is the most radical break, which means that a related plaintext can be recovered from a ciphertext meanwhile.

It has been proved that the subset sum problem is NP-Complete (NPC, shortly) [5], but when $D$ is less than 0.6463 and even 0.9408, a related plaintext can be recovered from a ciphertext through the procedures calling the $L^3$ algorithm, namely the $L^3$ algorithm [6][7][8], where

$$D = n / \log_2 \max_{1 \le i \le n}\{c_i\}$$
$$\approx \text{bit-size of a plaintext / bit-size of a ciphertext}$$

is called a sequence density or a knapsack density [6][7][8].

It is the $L^3$ algorithm that constructs a reduced basis of a lattice by the Gram-Schmidt orthogonal principle and seeks the shortest or an approximately shortest nonzero vector in the lattice [9]. In 1995, Schnorr and Hörner put forward a improved lattice basis reduction algorithm using pruned enumeration [10] which is able to crack some implementations of the Chor-Rivest cryptosystem with high densities [11] — a challenging scheme with $n = 103$ and $D = 1.271$ for example. In 1996, Ritter gave a new combination of the $L^3$ algorithm and pruned enumeration for the $l_\infty$-Norm shortest vector [12] through which the Orton public-key cryptosystem [13] involving compact knapsacks with $1 < D < 2$ was broken.

What is a compact knapsack?

When $a_i > (2^k - 1)\sum_{j=1}^{i-1} a_j$ for $i = 2, \ldots, n$, where $k \ge 2$ is an integer, $\{c_1, \ldots, c_n \mid c_i = pub(a_i)\}$ is called a compact knapsack. Correspondingly, the plaintext $x_1, \ldots, x_n \in [0, 2^k - 1]$, and the density $D = nk / \log_2 \max_{1 \le i \le n}\{c_i\}$.

In this paper, we propose a new public-key cryptosystem which is called REESSE2+, with $D \ge (n + 1) / 4$, and based on the lever function, an anomalous super-increasing sequence and an anomalous subset sum. It is not the successor of REESSE1+, but another application of the lever function; that we design REESSE2+ is not because REESSE1+ has any hidden security trouble, but because REESSE2+ has a smaller modulus and is more suitable for embedded or mobile CPUs.

Section II of the paper gives the definitions of an anomalous super-increasing sequence and an anomalous subset sum, and proves the two properties of an anomalous super-increasing sequence. Section III describes in detail the REESSE2+ scheme which contains 3 algorithms for a key pair, encryption, and decryption as well as owning the 5 characteristics, proves the correctness of decryption, and analyzes the nonuniqueness of a plaintext solution to a ciphertext in an extreme tiny probability.

Section IV discusses the necessity and sufficiency of the lever function, which does involve the the Shamir attack with $\ell(.)$ retrogressing to a constant, the ineffectiveness of the

Shenghui Su is with the College of Computers, Beijing University of Technology, Beijing 100124, P. R. China (e-mail: reesse@126.com).

Shuwang Lü is with the School of Graduate, Chinese Academy of Sciences, Beijing, 100039 P. R. China (e-mail: swlu@ustc.edu.cn).

minimum point method with $\ell(.)$ being injective, and the relation between the lever function and a random oracle. Section V expounds the security of REESSE2+ against extracting a private key from a public key, recovering a plaintext from a ciphertext plus a public key, and seeking a plaintext through meet -in-the-middle attacks or adaptive-chosen-ciphertext attacks, and argues that solving subset sums is restricted by NPC class, the length and density of a sequence, the density of a public key in REESSE2+ is in linear proportion to the length, and the cost of reducing ciphertexts via the $L^3$ lattice basis is not negligible.

Section VI analyzes the time complexities of the REESSE2+ encryption and decryption, and compares REESSE2+ in lengths and speeds with ECC and NTRU bearing matchable security, and points out the advantages of REESSE2+ over ECC and NTRU by data listed. ECC is the ElGamal analogue in an elliptic curve group [14].

Throughout the paper, we stipulate that $n \ge 120$, the sign '%' means 'modulo', 'gcd' represents the greatest common divisor, and 'log' denotes a logarithm to the base 2.

## II. An anomalous Super-increasing Sequence and an anomalous Subset Sum

***Definition 1**:* For $n$ positive integers $A_1, A_2, \ldots,$ and $A_n$, if every $A_i$ satisfies

$$A_i > \textstyle\sum_{j=1}^{i-1} (i-j)\, A_j,$$

where $i > 1$, then this series of integers is called an anomalous super-increasing sequence, denoted by $\{A_1, \ldots, A_n\}$, and shortly $\{A_i\}$.

***Definition 2:*** Assume that $b_1 \ldots b_n$ with every $b_i \in [0, 1]$ is a plaintext block, $\{X_1, \ldots, X_n\}$ is a sequence or set, and $E$ satisfies

$$E \equiv \textstyle\sum_{i=1}^{n} X_i b_i L_i,$$

where $L_i = \sum_{j=i}^{n} b_j$. Then $E$ is called an anomalous subset sum.

Notice that in definition 2, $\{X_1, \ldots, X_n\}$ is not required to be an anomalous super-increasing sequence.

Assume that the $i$-th bit of the plaintext block $b_1 \ldots b_n$ is 1 or 0, and all the bits after the $i$-th are 0. According to definition 1, obviously if $E > A_i$, then $b_i = 1$.

If the bit-string after the $i$-th bit contains $k$ 1-bits, and $E > (k + 1)A_i$, is there $b_i = 1$ yet?

***Property 1:*** Assume that $\{A_1, \ldots, A_n\}$ is an anomalous super-increasing sequence. Then, for $i > 1$ and any positive integer $k$, there exists $(k + 1)A_i > \sum_{j=1}^{i-1}(k + i - j)A_j$.

***Proof.***

Because $\{A_1, \ldots, A_n\}$ is an anomalous super-increasing sequence, we have $A_i > \sum_{j=1}^{i-1}(i-j)A_j$ for $i = 2, \ldots, n$. That is,

$$A_i > (i-1)A_1 + (i-2)A_2 + \ldots + A_{i-1}. \quad (1)$$

It is easily inferred from (1) that

$$A_i > A_1 + A_2 + \ldots + A_{i-1}. \quad (2)$$

Multiplying either side of (2) by $k$ makes

$$kA_i > kA_1 + kA_2 + \ldots + kA_{i-1}. \quad (3)$$

Adding (3) to (1) yields

$$kA_i + A_i > (k+i-1)A_1 + (k+i-2)A_2 + \ldots + (k+1)A_{i-1},$$

which is written shortly as $(k+1)A_i > \sum_{j=1}^{i-1}(k+i-j)A_j$. □

By property 1, when the anomalous subset sum $E > (k + 1)A_i$, there is $b_i = 1$.

***Property 2:*** For any positive integer $m \le n$, if randomly select $m$ elements from the anomalous super-increasing sequence $\{A_i\}$ and construct a subset $\{Ax_1, \ldots, Ax_m\}$ in original order, the anomalous subset sum $E = mAx_1 + (m-1)Ax_2 + \ldots + Ax_m$ is uniquely determined, that is, the mapping from $E$ to $\{Ax_1, \ldots, Ax_m\}$ is one-to-one.

***Proof.*** By contradiction.

Presume that $E$ is acquired from two different incompatible subsequences $\{Ax_1, \ldots, Ax_m\}$ and $\{Ay_1, \ldots, Ay_h\}$. Incompatibility means that the set

$$\{Ax_1,\ldots,Ax_m\} \not\subset \{Ay_1,\ldots,Ay_h\} \text{ and } \{Ay_1,\ldots,Ay_h\} \not\subset \{Ax_1,\ldots,Ax_m\}.$$

Then,

$$E = mAx_1 + (m-1)Ax_2 + \ldots + Ax_m$$
$$= hAy_1 + (h-1)Ay_2 + \ldots + Ay_h.$$

First, observe $Ax_m$ and $Ay_h$. If $Ax_m = Ay_h$, continue to observe $Ax_{m-1}$ and $Ay_{h-1}$.

Without loss of generality, let $Ax_{m-k} \ne Ay_{h-k}$ and the subscript $x_{m-k} > y_{h-k}$. In terms of definition 1 and property 1, we have

$$(k+1)Ax_{m-k} > hAy_1 + (h-1)Ay_2 + \ldots + (k+1)Ay_{h-k},$$

which indicates that

$$mAx_1 + (m-1)Ax_2 + \ldots + Ax_m \ne hAy_1 + (h-1)Ay_2 + \ldots + Ay_h.$$

It is in direct contradiction to the presumption.

Therefore, the mapping relation between $E$ and $\{Ax_1, \ldots, Ax_m\}$ is one-to-one. □

***Definition 3**:* In a public key cryptosystem, the parameter $\ell(i)$ in the key transform is called the lever function, if it has the following features [15]:

- $\ell(.)$ is an injection from integers to integers, its domain is $[1, n]$, and codomain $[1, M)$. Let $Ł_n$ represent the collection of all injections from the domain to the codomain, then $\ell(.) \in Ł_n$ and $|Ł_n| \ge \mathrm{A}_n^n = n(n-1) \ldots 1$.
- The mapping between $i$ and $\ell(i)$ is established randomly without an analytical formula, so every time a public key is generated, the function $\ell(.)$ is distinct.
- There does not exist any dominant or special mapping from $\ell(.)$ to a public key.
- An attacker have to consider all the arrangements of the sequence $\{\ell(i) \mid i = 1, \ldots, n\}$ when extracting a related private key from a public key. Thus, if $n$ is large enough, it is infeasible for the attacker to search the arrangements exhaustively.
- A receiver owning a private key only needs to consider the linear or square accumulative sum of the sequence $\{\ell(i)\}$ when recovering a related plaintext from a ciphertext. Thus the time complexity of decryption is polynomial in $n$, and the decryption is feasible.

Obviously, there is the large amount of calculation on $\ell(.)$ at 'a public terminal', and the small amount of calculation on $\ell(.)$ at 'a private terminal'.

## III. Design of the REESSE2+ Encryption Scheme

### A. The Key Generation Algorithm

This algorithm is employed by a third-party authority. Every user is given a pair of keys.

S1: Randomly generate an anomalous super-increasing sequence $\{A_1, \ldots, A_n\}$ with every $A_i$ being even.

S2: Find an integer $M$ making $M > \sum_{i=1}^{n}(n+1-i)A_i$ and $1.585n \le \log M \le 2n$.

S3: Pick integers $W, Z < M$ meeting $\gcd(W, M) = 1$ and $M / \gcd(M, Z) \approx n^3 2^{n/2}$.
Calculate $W^{-1}$ by $WW^{-1} \equiv 1\ (\% M)$ and $-Z$ by $Z + (-Z) \equiv 0\ (\% M)$.

S4: Produce pairwise distinct values $\ell(1), \ldots, \ell(n) \in \Omega = \{5, \ldots, n+4\}$ at will.

S5: Compute the sequence $\{C_1, \ldots, C_n \mid C_i \leftarrow (A_i + Z\ell(i))W \,\%\, M\}$.

At last, the public key is $(\{C_i\}, M)$, the private key $(\{A_i\}, W^{-1}, -Z, M)$, and $\{\ell(i)\}$ discarded.

Clearly, when $Z = 0$, the REESSE2+ transform retrogresses to the MH transform.

Letting $\Omega = \{5, \ldots, n+4\}$ is to keep conformity with the REESSE1+ cryptosystem [15]. In fact, $\Omega$ may take other values — $\Omega = \{1, \ldots, n\}$ for example. The principles for selecting $\Omega$ are that 1) $\ell(i) \ge 1$; 2) the elements of $\Omega$ are pairwise distinct; 3) decryption time complexity does not exceed $O(n^3)$ arithmetic steps.

When we generate $\{A_1, \ldots, A_n\}$, let every $A_i$ be slightly greater than $\sum_{j=1}^{i-1}(i-j)A_j$ for $i > 2$ lest the length of $\log M$ should be too larger. In this way, $A_i > \sum_{j=1}^{i-1}(i-j)A_j$ may be regarded as

$$A_i \approx \sum_{j=1}^{i-1}(i-j)A_j.$$

Further,

$$\begin{aligned} A_{i+1} &> \sum_{j=1}^{i}(i+1-j)A_j \\ &= iA_1 + (i-1)A_2 + \ldots + A_i \\ &= ((i-1)A_1 + (i-2)A_2 + \ldots + A_{i-1}) + \\ &\quad (A_1 + A_2 + \ldots + A_{i-1}) + A_i \\ &\approx A_i + A_i + (A_1 + A_2 + \ldots + A_{i-1}). \end{aligned}$$

Similarly, since slightly $A_{i+1} > \sum_{j=1}^{i}(i+1-j)A_j$, it is not difficult to understand that

$$A_{i+1} \approx A_i + A_i + (A_1 + A_2 + \ldots + A_{i-1}).$$

Therefore, when $i > 2$, and slightly $A_i > \sum_{j=1}^{i-1}(i-j)A_j$, we see

$$2A_i < A_{i+1} \le 3A_i.$$

Namely,

$$2 < A_{i+1} / A_i \le 3.$$

As two exceptions, when $i = 1$, $1 < A_{i+1} / A_i \le 3$, and when $i = 2$, $2 \le A_{i+1} / A_i \le 3$. For example, there are the sequences $\{2, 3, 8, \ldots\}$, and $\{2, 5, 10, \ldots\}$.

Moreover, thanks to the existence of $Z$, $A_1$ is allowed to take a small value.

For example, let $n = 6$ and $\{A_1, \ldots, A_6\} = \{1, 2, 5, 13, 34, 89\}$, and then $A_6 / A_5 \approx 2.6176$, $A_5 / A_4 \approx 2.6153$, $A_4 / A_3 \approx 2.6$, and $A_3 / A_2 \approx 2.5$.

Additionally, $\{3^0, 3^1, \ldots, 3^{n-1}\}$ with $M = 3^n$ is evidently an anomalous super-increasing sequence, and there is $\log M = \log 3^n \approx \log 2^{1.585n} \approx 1.585n$, which indicates that the condition $\log M \le 2n$ at step 2 is easily achieved. Meanwhile, we observe that when $2 < A_{i+1} / A_i \le 3$ and $\log M \le 2n$, the number of anomalous super-increasing sequences will be greater than $3^n$ because the difference between possible minimal $A_i$ and possible maximal $A_i$ is not less than 2 for $i > 3$, and also observe that when $2 < A_{i+1} / A_i \le 3$, $\log M \le 2n$, possible maximal $A_1$ is roughly $3^m$, where the exponential $m$ satisfies $1.585(n+m) \approx 2n$. For instance, when $n = 120$ and $\log M = 2n$, $m \approx 31$.

### B. The Encryption Algorithm

Assume that $(\{C_i\}, M)$ is the public key, and $b_1 \ldots b_n$ is an $n$-bit plaintext block or symmetric key, which may possibly be padded with a random binary string.

S1: Let $\bar{E} \leftarrow 0$, $L \leftarrow 0$, $i \leftarrow n$.

S2: If $b_i = 1$, do $L \leftarrow L + 1$ and $\bar{E} \leftarrow \bar{E} + LC_i \,\%\, M$.

S3: Set $i \leftarrow i - 1$.
If $i \ge 1$, go to S2, or else end.

After the algorithm is executed, the ciphertext $\bar{E}$ is gained.

Apparently, $\bar{E}$ may be expressed as $\bar{E} \equiv \sum_{i=1}^{n} C_i b_i L_i\ (\% M)$, where $L_i = \sum_{j=i}^{n} b_j$. In terms of definition 2, $\bar{E}$ is an anomalous subset sum.

### C. The Decryption Algorithm

Assume that $(\{A_i\}, W^{-1}, -Z, M)$ is the private key, and $\bar{E}$ is the ciphertext.

In advance, compute

$$E_k = \sum_{i=1}^{k}(k+1-i)A_i \text{ for } k = n, \ldots, 1,$$

$$\dot{E}_k = \sum_{i=1}^{k}A_i \text{ for } k = n-1, \ldots, 1,$$

and store individually $E_n, \ldots, E_1$, $\dot{E}_{n-1}, \ldots, \dot{E}_1$ to a data segment of a decryption program.

S1: Compute $\bar{E} \leftarrow \bar{E}W^{-1} \,\%\, M$.

S2: Repeat $\bar{E} \leftarrow \bar{E} + (-Z) \,\%\, M$ until $\bar{E}$ is even and $\bar{E} \le E_n$.

S3: Let $b_1 \ldots b_n \leftarrow 0$, $E \leftarrow \bar{E}$, $L \leftarrow 0$, $i \leftarrow n$.

S4: If $E \ge (L+1)A_i$, do $b_i \leftarrow 1$, $L \leftarrow L + 1$ and $E \leftarrow E - LA_i$.

S5: Set $i \leftarrow i - 1$.
If $i \ge 1$ and $0 < E \le E_i + L\dot{E}_i$, go to S4.

S6: If $E \ne 0$, go to S2, or else end.

At last, the $b_1 \ldots b_n$ is the original plaintext block or the symmetric key.

This algorithm can always terminate normally as long as $\bar{E}$ is a true ciphertext.

### D. Correctness of the Decryption Algorithm

Because $(\mathbb{Z}_M, +)$ is an Abelian, namely commutative group, $\forall k \in [0, M)$, there is

$$kZ + k(-Z) \equiv kZ + (-kZ) \equiv 0\ (\% M).$$

Let $b_1 \ldots b_n$ be an $n$-bit plaintext, and $k = (\sum_{i=1}^{n} \ell(i) b_i L_i)$, where $L_i = \sum_{j=i}^{n} b_j$.

We need to prove that $\bar{E} W^{-1} + k(-Z) \equiv \sum_{i=1}^{n} A_i b_i L_i \equiv E\ (\% M)$.

According to section III.B, $\bar{E} \equiv \sum_{i=1}^{n} C_i b_i L_i\ (\% M)$, where $C_i \equiv (A_i + Z\ell(i))W\ (\% M)$, hence

$$\begin{aligned} \bar{E} W^{-1} + k(-Z) &\equiv (\textstyle\sum_{i=1}^{n} C_i b_i L_i) W^{-1} + k(-Z) \\ &\equiv W^{-1} \textstyle\sum_{i=1}^{n} (A_i + Z\ell(i)) W b_i L_i + k(-Z) \end{aligned}$$

$$\equiv \sum_{i=1}^{n}(A_i\, b_i\, L_i + Z\ell(i)\; b_i L_i) + k(\text{-}Z)$$
$$\equiv \sum_{i=1}^{n} A_i b_i L_i + \sum_{i=1}^{n} Z\ell(i)\, b_i L_i + k(\text{-}Z)$$
$$\equiv E + Z\,k + (\text{-}k\,Z)$$
$$\equiv E\ (\%\ M).$$

Apparently, the above proof gives a method for seeking $E$.

Notice that in practice, the plaintext $b_1 \ldots b_n$ is unknowable in advance, so we have no way to directly compute $k$. However, because the range of $k \in [1, \sum_{i=1}^{n} i^2]$ is very narrow, we may search $k$ heuristically by adding $(\text{-}Z)\ \%\ M$, and verify whether $E$ is equal to 0 after some items $(A_i L_i)$ are subtracted from $E$. It is known from section III.C that the original plaintext $b_1 \ldots b_n$ is acquired at the same time the condition $E = 0$ is satisfied.

### *E. Uniqueness of a Plaintext Solution to a Ciphertext*

Because $\{C_i\}$ is not an anomalous super-increasing sequence, the mapping from the subsequence $\{C_{x_1}, \ldots, C_{x_m}\}$, which is in original order, to the anomalous subset sum $\bar{E}$ is theoretically many-to-one. It might possibly result in the nonuniqueness of the plaintext solution $b_1 \ldots b_n$ when $\bar{E}$ is being unveiled.

Suppose that the ciphertext $\bar{E}$ can be obtained from two different subsequences, that is,

$$\bar{E} \equiv m\,C_{x_1} + (m-1)C_{x_2} + \ldots + C_{x_m}$$
$$\equiv h\,C_{y_1} + (h-1)C_{y_2} + \ldots + C_{y_h}\ (\%\ M).$$

Then,

$$(m(A_{x_1} + Z\ell(x_1)) + (m-1)(A_{x_2} + Z\ell(x_2)) + \ldots + (A_{x_m} + Z\ell(x_m)))W$$
$$\equiv (h(A_{y_1} + Z\ell(y_1)) + (h-1)(A_{y_2} + Z\ell(y_2)) + \ldots + (A_{y_h} + Z\ell(y_h)))W$$
$$(\%\ M).$$

It follows that

$$(mA_{x_1} + (m-1)A_{x_2} + \ldots + A_{x_m} + Zk_1)$$
$$\equiv (hA_{y_1} + (h-1)A_{y_2} + \ldots + A_{y_h} + Zk_2)\ (\%\ M),$$

where

$$k_1 = m\ell(x_1) + (m-1)\ell(x_2) + \ldots + \ell(x_m), \text{ and}$$
$$k_2 = h\ell(y_1) + (h-1)\ell(y_2) + \ldots + \ell(y_h).$$

Without loss of generality, let $k_1 \geq k_2$. Because $(\mathbb{Z}_M, +)$ is an Abelian group, there is

$$Z(k_1 - k_2)$$
$$\equiv (hA_{y_1} + (h-1)A_{y_2} + \ldots + A_{y_h}) - (mA_{x_1} + (m-1)A_{x_2} + \ldots + A_{x_m})\ (\%\ M),$$

which is written shortly as

$$Z(k_1 - k_2) \equiv \sum_{i=1}^{h}(h+1-i)A_{y_i} + \sum_{j=1}^{m}(m+1-j)(\text{-}A_{x_j})\ (\%\ M).$$

Letting

$$\theta \equiv \sum_{i=1}^{h}(h+1-i)A_{y_i} + \sum_{j=1}^{m}(m+1-j)(\text{-}A_{x_j})\ (\%\ M)$$

gives

$$Z(k_1 - k_2) \equiv \theta\ (\%\ M).$$

In a cryptographic application, $Z$ is fixed. If the value of the variable $(k_1 - k_2)$ exists, the condition $\gcd(Z, M) \mid \theta$ must be satisfied [16]. Thus, we need to observe the probability that the condition holds.

If an adversary tries to attack an 80-bit symmetric key or plaintext block through the exhaustive search, and a computer can verify trillion values per second, it will take 38334 years to verify up all the potential values, which indicates that currently 80 bits are quite enough for the security of a symmetric key or plaintext block.

Let $\tilde{n}$ denote the number of values formed from $\sum_{i=1}^{h}(h+1-i)$ $A_{y_i} + \sum_{j=1}^{m}(m+1-j)(\text{-}A_{x_j})\ (\%\ M)$.

If the coefficients $(h+1-i)$ and $(m+1-j)$ before $A_{y_i}$ and $(\text{-}A_{x_j})$ are neglected, one of $A_t$, $\text{-}A_t$ and null, where $t \in [1, n]$, may occurs in $\sum_{i=1}^{h} A_{y_i} + \sum_{j=1}^{m} \text{-}A_{x_j}\ \%\ M$ ($A_t$ and $\text{-}A_t$ will counteract each other if both do in the expression at the same time). Therefore, on this assumption, $\tilde{n}$ is less than $3^n \approx 2^{1.585n}$ exclusive of the repeated values. If the coefficients before $A_{y_i}$ and $\text{-}A_{x_j}$ are considered, the number of values formed from $\sum_{i=1}^{h}(h+1-i)A_{y_i}$ is $2^n$ while the number of values formed from $\sum_{j=1}^{m}(m+1-j)(\text{-}A_{x_j})$ is at most $2^n$ since $\{\text{-}A_1, \ldots, \text{-}A_n\}$ is not necessarily an anomalous super-increasing sequence, which means that $\tilde{n}$ is at most the maximum between $2^{2n}$ and $M - 1$.

It is known from section III.A that $M / \gcd(M, Z) \approx n^3 2^{n/2}$ holds, which manifests that there are $n^3 2^{n/2}$ integers which can be divided exactly by $\gcd(M, Z)$, and distribute uniformly on the interval $[1, M]$. Obviously, the probability that any arbitrary integer $\in (1, M)$ can be divided exactly by $\gcd(M, Z)$ is $n^3 2^{n/2} / M$, where the numerator is not less than $2^{80}$ when $n \geq 120$.

Suppose that $\log M = 1.585n$, namely $M \approx 2^{1.585n}$, and the values of $\theta$ distribute uniformly on the interval $[1, M)$. Then, the probability of $\gcd(Z, M) \mid \theta$ is

$$(\tilde{n}(n^3 2^{n/2} / M)) / \tilde{n} \approx n^3 2^{n/2} / 2^{1.585n} \leq n^3 / 2^n.$$

If the values of $\theta$ do not distribute uniformly on the interval $[1, M)$, then the probability of $\gcd(Z, M) \mid \theta$ will be much less than $n^3 / 2^n$.

Notice that if

$$mC_{x_1} + (m-1)C_{x_2} + \ldots + C_{x_m} \equiv L_{y_1}C_{y_1} + L_{y_2}C_{y_2} + \ldots + L_{y_h}C_{y_h}\ (\%\ M)$$

with $L_{y_i} + 1 \neq L_{y_{i-1}}$, then the right of the equation will not influence the uniqueness of a decrypted plaintext.

The preceding analysis makes it clear that when $\log M \geq 1.585n$ and $M / \gcd(M, Z) \approx n^3 2^{n/2}$, the probability that the plaintext solution $b_1 \ldots b_n$ is not unique is less than $n^3 / 2^n$, and almost zeroth when $n \geq 120$. Thus, the decryption algorithm can always recover the original plaintext from the ciphertext $\bar{E}$, which is also validated by the program in C language.

### *F. Characteristics of REESSE2+*

REESSE2+ owes the following characteristics compared with classical MH, RSA and ElGamal cryptosystems.

- The key transform $C_i \equiv (A_i + Z\ell(i))W\ (\%\ M)$ is a compound function, and contains four independent variables. That is, the $n$ equations contain $2n + 2$ unknown variables. Hence, REESSE2+ is a multivariate cryptosystem [15].
- If any of $A_i$, $Z$, $W$ and $\ell(i)$ is determined, the relation among the three remainders is still nonlinear — thus there is very complicated nonlinear relations among $A_i$, $Z$, $W$ and $\ell(i)$.
- There is indeterminacy of $\ell(i)$. On condition that $C_i$, $Z$ and $W$ are determined, $A_i$ and $\ell(i)$ can not be determined, and even have no one-to-one relation for $\gcd(Z, M) > 1$. On condition that $C_i$, $W$ and $A_i$ are determined, $Z$ and $\ell(i)$ can not be determined, and also have no one-to-one relation when $\gcd(\ell(i), M) > 1$.
- There is insufficiency of the key mapping. Roundly speaking, a private key in REESSE2+ includes $\{A_i\}$, $Z$, $W$ and

discarded $\{\ell(i)\}$ four parts, but there is only a dominant mapping from $\{A_i\}$ to $\{C_i\}$. Thereby, the reversibility of the function is not obvious, and inferring a private key is intractable resorting to mathematical methods.

- Since the elements of the set $\Omega$ are not fixed — $\Omega = \{\delta + i \% M \mid i = 5, \ldots, n + 4\}$ for example, and the diversity of the mathematical definition of an anomalous subset sum, REESSE2+ is a sort of flexible public key cryptosystem.

## IV. Necessity and Sufficiency of the Lever Function

The necessity of the lever function $\ell(.)$ means that if the private key in REESSE2+ is secure, $\ell(.)$ as a function must exist in the key transform. The equivalent contrapositive assertion is that if $\ell(.)$ does not exist or is a constant in the key transform, the private key in REESSE2+ will be insecure.

The sufficiency of the lever function $\ell(.)$ means that if $\ell(.)$ as a injection exists in the key transform, the private key in REESSE2+ is secure, that is, the Shamir attack method based on the accumulation of minimum points is ineffective.

### A. Potential Attack by the Shamir Method When $\ell(.) = k$ Forever

If a private key is insecure, a plaintext must be insecure. Hence, the security of the private key is fundamental and all-around.

It is known from section III.A that when $\ell(.)$ is a constant $k$, the key transform becomes as $C_i \equiv (A_i + Zk)W$ (% $M$) which is equivalent to $\ell(.)$ being ineffectual or nonexistent. Notice that the variation on $\ell(.)$ does not influence the correctness of all the algorithms.

When $C_i \equiv (A_i + Zk)W$ (% $M$), there are

$$C_1 \equiv (A_1 + Zk)W \ (\% \ M),$$
$$C_2 \equiv (A_2 + Zk)W \ (\% \ M),$$
$$\vdots$$
$$C_n \equiv (A_n + Zk)W \ (\% \ M).$$

Subtracting $C_1$ from every equality beginning with $C_2$ gives a difference sequence

$$\{C_2 - C_1, C_3 - C_1, \ldots, C_n - C_1\}.$$

That is to say, $C_i - C_1 \equiv (A_i - A_1)W$ (% $M$) for $i = 2, \ldots, n$. To a greater degree, $A_i - A_1 \equiv (C_i - C_1)W^{-1}$ (% $M$).

Notice that because $W$ is unknown, and $A_i$ is likely large with relation to $M$, the continued fraction analysis [15] is ineffectual on the REESSE2+ with $\ell(.) = k$.

Section III.A tells us that when $\{A_1, \ldots, A_n\}$ is an anomalous super-increasing sequence, and satisfies the constraint conditions, there is a rough proportion between $A_{i+1}$ and $A_i$, namely $2 < A_{i+1} / A_i \le 3$ for every $i > 2$. Considering that $A_1$ is a small integer, there also exists a rough proportion between $(A_{i+1} - A_1)$ and $(A_i - A_1)$ when the subscript $i$ is comparatively large. Therefore, in light of Shamir extremum method [4], $W$ and $\{A_2 - A_1, A_3 - A_1, \ldots, A_n - A_1\}$ might be found out. Additionally, in consideration of $A_1$ being small, $A_1$ can be guessed in acceptable time, which indicates $\{A_1, \ldots, A_n\}$ may be possibly inferred. It follows that $(Zk)$ can be figured out.

The above analysis illustrates that when the lever function $\ell(.)$ is the constant $k$, a related private key is likely deduced from a public key and further a related plaintext is likely recovered from a ciphertext. Thus, $\ell(.)$ as an injective function is necessary to the REESSE2+ scheme.

### B. Ineffectiveness of the Minimum Point Attack with $\ell(.)$ being Injective

In the MH knapsack cryptosystem, every item of the super-increasing sequence $\{a_1, \ldots, a_n\}$ is approximately double the previous item of itself, that is,

$$a_n < 2^{-1}M, a_{n-1} < 2^{-2}M, \ldots, a_1 < 2^{-n}M.$$

Thus, Shamir broke the MH system by utilizing this regularity.

In the REESSE2+ scheme, let $A_i + Z\ell(i) \equiv C_iW^{-1}$ (% $M$), and then

$$\{A_1 + Z\ell(1), A_2 + Z\ell(2), \ldots, A_n + Z\ell(n)\}$$

is not a super-increasing sequence or an anomalous super-increasing sequence, but a stochastic sequence.

Because there does not exist an fixed proportional relation between $(A_{i+1} + Z\ell(i + 1))$ and $(A_i + Z\ell(i))$, namely the value

$$(A_{i+1} + Z\ell(i + 1)) / (A_i + Z\ell(i))$$

is not approximately unvarying, the minimum point attack of Shamir is ineffectual on the sequence $\{A_1 + Z\ell(1), A_2 + Z\ell(2), \ldots, A_n + Z\ell(n)\}$, which indicates the lever function is sufficient for the security of a private key against the Shamir attack method.

### C. Relation between $\ell(.)$ and a Random Oracle

If an adversary tries to find $i$, $j$ and $k$ such that $\ell(k) = \ell(i) + \ell(j)$, where $i < j < k$, he will be confronted with the two difficulties:

- Due to $W \in [1, M)$, the discriminant
  $$A_i + A_j - A_k \equiv (C_i + C_j - C_k)W^{-1} \ (\% \ M)$$
  can not be verified in polynomial time.
- The function $\ell(.)$ bears indeterminacy. For example, when $\ell(i) + \ell(j) \ne \ell(k)$, there exist
  $$C_i \equiv (A'_i + Z'\ell'(i))W' \ (\% \ M),$$
  $$C_j \equiv (A'_j + Z'\ell'(j))W' \ (\% \ M),$$
  $$C_k \equiv (A'_k + Z'\ell'(k))W' \ (\% \ M)$$
  such that $\ell'(i) + \ell'(j) = \ell'(k)$, $A'_i < A'_j$, and $2A'_i + A'_j < A'_k$.

In what follows, the indeterminacy of $\ell(.)$ will be explained further.

A function or algorithm is randomized if its output depends not only on the input but also on some random ingredient, namely if its output is not uniquely determined by the input. In other words, a random function or algorithm outputs the different value every time it receives the same input. According to this definition, the randomness of a function or algorithm is almost equivalent to the indeterminacy.

Of course, a random function or algorithm may be a random oracle which is a theoretical black box, and answers to every query with a completely random and unpredictable value chosen uniformly from its output domain [17][18].

Suppose that $R_\ell$ is an random oracle for the lever function value $\{\ell(i)\}$.

We construct $R_\ell$ as follows:

Input: $\{C_1, \ldots, C_n\}$, $M$.

Output: $\{\ell(1), \ldots, \ell(n)\}$.

S1: Randomly produce an anomalous super-increasing sequence $\{A_1, \ldots, A_n\}$ such that $\sum_{i=1}^{n}(n+1-i)A_i < M$, where every $A_i$ is even.

S2: Pick integers $W, Z < M$ such that $\gcd(W, M) = 1$ and $M / \gcd(Z, M) \approx n^3 2^{n/2}$.

S3: Compute $\ell(i)$ by $C_i \equiv (A_i + Z\ell(i))W$ (% $M$) if $\gcd(Z, M) \mid (C_i W^{-1} - A_i)$ for $i = 1, \ldots, n$.

S4: If every $\ell(i)$ is already computed, return $\{\ell(1), \ldots, \ell(n)\}$; or else go to S1.

According to definition 2, every $\ell(i)$ may be outside of $[5, n + 4]$ and pairwise inconsecutive, namely any $\ell(i) \in [1, M-1]$ is up to the definition and the requirement. By the way, $\{A_i\}$, $W$ and $Z$ as side results may be outputted.

The above algorithm illustrates that the output $\{\ell(i)\}$ depends not only on $\{C_i\}$ and $M$ but also on random $W$, $Z$ and $\{A_i\}$. Namely every time for the same input $(\{C_i\}, M)$, the output $\{\ell(i)\}$ is different or randomized. Therefore, $R_\ell$ is exactly a random oracle, and it is impossible that through $R_\ell$ the adversary obtains the specific $\{\ell(i)\}$ and other private key part generated by the key algorithm. Additionally, because the above algorithm is a random oracle, we do not need to be concerned for its running time.

The discussion in the section manifests soundly that any indeterministic reasoning, if it exists, is ineffectual on the private key in REESSE2+.

## V. Security Analysis of the REESSE2+ Scheme

### A. Extracting a Private Key from a Public Key Is Intractable

A public key may be treated as the special cipher of a related private key. A ciphertext is the integration of a public key and a plaintext, so the ciphertext has no direct help to inferring the private key. In this section, we make further analysis of the security of the private key in terms of exhaustive search.

In the REESSE2+ scheme, the key transform is $C_i \equiv (A_i + Z\ell(i))W$ (% $M$), where $\ell(i) \in \{5, \ldots, n+4\}$, $1.585n \le \log M \le 2n$, $W$ meets $\gcd(W, M) = 1$, and $Z$ meets $M / \gcd(M, Z) \approx n^3 2^{n/2}$.

If an adversary attempt to guess $W$, because the number of potential values of $W$ equals $\varphi(M)$, the probability of hitting $W$ is $1 / \varphi(M)$. Clearly, we can make $\varphi(M) \ge n^3 2^{n/2}$ by taking a fit $M$.

Because of $\log M \le 2n$, $M$ can be factorized in tolerable time, which indicates that $\gcd(M, Z)$ may possibly is found out in polynomial time. Furthermore, $Z$ may be guessed. However, owing to $M/\gcd(M, Z) \approx n^3 2^{n/2}$, the probability of successfully guessing $Z$ by brute force is approximately $1 / (n^3 2^{n/2})$, not greater than $1 / 2^{80}$ as $n \ge 120$. Clearly, it is almost zeroth.

If the adversary guess the sequence $\{\ell(1), \ldots, \ell(n)\}$, namely an arrangement of $\{5, \ldots, n+4\}$, the probability of successfully guessing $\{\ell(1), \ldots, \ell(n)\}$ is $1 / n!$, where $n! = n(n-1)\ldots1$ is the factorial of $n$.

According to section III.A, the number of all possible $\{A_1, \ldots, A_n\}$ is greater than $3^n$ when $2 < A_{i+1} / A_i \le 3$ and $\log M \le 2n$. Therefore, the probability of successfully guessing $\{A_i\}$ by brute force is $1 / 3^n$.

If the adversary assume the values of $W$, $Z$ and $\{\ell(i)\}$, and $\{A_i\}$ computed from $A_i \equiv W^{-1} C_i - Z\ell(i)$ (% $M$) is an anomalous super-increasing sequence, the guess is successful. However, the time complexity of such a guess is up to $O(n^3 2^{n/2} n! \varphi(M))$.

If assume the values of $Z$, $\{A_i\}$ and $\{\ell(i)\}$, the sufficient and necessary condition for $(A_i + Z\ell(i))W \equiv C_i$ (% $M$) to have solutions is $\gcd(A_i + Z\ell(i), M) \mid C_i$. Hence, $W$ does not always exist. Similarly, $Z$ does not always exist, and neither does $\ell(i)$.

### B. Recovering a Plaintext from a Ciphertext and a Public Key Is Intractable

*1) Solving Subset Sums Is Restricted by NPC, Lengths and Densities:*

The subset sum problem is similar to the partition problem and the integer programming problem [5][19]. If the solution to a subset sum is just the shortest vector in a related lattice, and it could be sought in polynomial time, the subset sum problem will degenerate from the NP-Complete class.

Coster, Joux, LaMacchia etc showed that when the density $D < 0.9408$, the solution vector to a normal subset sum is the shortest [8]. Schnorr and Hörner showed that for the Chor-Rivest cryptosystem with $D < 1.271$, the solution vector to a related subset sum is the shortest [10]. Ritter showed that for the Orton cryptosystem with $1 < D < 2$, the solution vector to a related subset sum is the shortest in $l_\infty$-Norm [12].

The general $l_\infty$-Norm shortest vector problem is known to be NP-Complete [12][20], and thus when the length $n$ augments, even though the density is still kept reasonable, any breaking method through $l_\infty$-Norm, including the Ritter method, will gradually disable. The $l_2$-Norm shortest vector problem is open [8][12] for a fixed dimension which equals $n + 1$, and NP-hard for a varying dimension [21]. Thus, we conjecture that when $n$ increases, and $D$ is kept unvaried, the probability of breaking subset sum ciphertexts through $l_2$-Norm [8][22] will gradually decrease, which is validated by our experiments afterwards.

Of course, when the density $D$ increases, and $n$ is kept unvaried, the probability of solving subset sums by the same algorithm will also decrease.

Coster, Joux, LaMacchia etc thought that an $l_\infty$-Norm lattice oracle yields a better density bound than an $l_2$-Norm lattice oracle [8], which seems to be validated by [12]. They also judged that "we cannot hope to asymptotically improve the 0.9408 bound by reducing a polynomial number of bases with different $b_{n+1}$ vectors. However for small dimensions it might be possible to improve the bound even though any such advantage will disappear as $n$ grows." [8], which is likewise applicable to the Ritter method since the main difference is barely in standards for vector measure between the two methods.

The foregoing argumentation makes it clear that so long as the density $D$ of a sequence is greater than 2, and the length $n$ is large enough — $n \ge 120$ for example, a subset sum ciphertext will be secure because in this case, the shortest nonzero vector in a lattice either can not be transformed into the right solution to a subset sum, or can not be found out in polynomial time.

Notice that in general, when the sequence density $D > 1$, there will be many subsets of weights with the same sum, and thus a sequence with $D > 1$ applied to a cryptosystem should be devised elaborately.

*2) Density of a Public Key in REESSE2+ Is in Linear Proportion to n:*

In the REESSE2+ scheme, a ciphertext $\bar{E}$ is one anomalous subset sum, namely $\bar{E} \equiv \sum_{i=1}^{n} C_i b_i L_i$ (% $M$), where $L_i = \sum_{j=i}^{n} b_j$, and $\sum_{i=1}^{n} b_i L_i \le n(n+1)/2$.

To seek a solution $(b_1 L_1, \ldots, b_n L_n)$ to $\bar{E}$ in polynomial time, an adversary must construct a suitable lattice basis according to $\{C_1, \ldots, C_n\}$. Clearly, an intuitionistic way is to let the lattice $\mathbb{L}$ be spanned by the linearly independent vectors

$$\vec{b}_1 = (1, 0, \ldots, 0, NC_1),$$
$$\vec{b}_2 = (0, 1, \ldots, 0, NC_2),$$
$$\vdots$$
$$\vec{b}_n = (0, 0, \ldots, 1, NC_n),$$
$$\vec{b}_{n+1} = (0, 0, \ldots, 0, NS),$$

where $N > (n^{1/2}) / 2$ is a positive integer, $S \in \{\bar{E} + 0M, \bar{E} + 1M, \ldots, \bar{E} + (n(n+1)/2)M\}$, and the dimension of each $\vec{b}_i$ is $n + 1$. Because all the nonzero elements of a solution vector are distinct from one another, and greater than or equal to 1, $\vec{b}_{n+1}$ may not be designed as (½, ½, …, ½, $NS$).

However, it is specious — when the elements of a solution vector distribute on a interval beyond [0, 1], even if the density $D < 0.6463$, the shortest vector in a lattice is not the solution to an anomalous subset sum.

For example, let {211, 122, 300} be a sequence with the density $D = 3 / 9 < 0.6463$, and a related Diophantine equation be $211x + 122y + 300z = 1177$ with a constraint that the nonzero items of $\{x, y, z\}$ descend gradually by 1. It is easily understood that the uniquely fit solution is $(x, y, z) = (3, 2, 1)$.

By the basis reduction, let the lattice $\mathbb{L}' = \mathbb{Z}\vec{b}_1 + \mathbb{Z}\vec{b}_2 + \mathbb{Z}\vec{b}_3 + \mathbb{Z}\vec{b}_4$, where $\mathbb{Z}$ is the integer set, and

$$\vec{b}_1 = (1, 0, 0, 211N),$$
$$\vec{b}_2 = (0, 1, 0, 122N),$$
$$\vec{b}_3 = (0, 0, 1, 300N),$$
$$\vec{b}_4 = (0, 0, 0, 1177N).$$

Then, $3\vec{b}_1 + 2\vec{b}_2 + 1\vec{b}_3 - 1\vec{b}_4 = (3, 2, 1, 0)$ is a solution vector, and $2\vec{b}_1 - 1\vec{b}_2 - 1\vec{b}_3 - 0\vec{b}_4 = (2, -1, -1, 0)$ is also a solution vector. The former is satisfied with the constraint, but its distance $(3^2 + 2^2 + 1^2 + 0^2)^{1/2} > (2^2 + (-1)^2 + (-1)^2 + 0^2)^{1/2}$ in $l_2$-Norm, or $3 > 2$ in $l_\infty$-Norm. That is, the solution (3, 2, 1, 0) is not the shortest nonzero vector in $\mathbb{L}'$.

Therefore, the adversary has to transform an anomalous subset sum to a normal subset sum to obtain a right solution vector each element of which is either 0 or 1, and further he need to consider the new layout of the sequence $\{C_1, \ldots, C_n\}$ as a public key.

In conformity with $\bar{E} \equiv \sum_{i=1}^{n} C_i b_i L_i$ (mod $M$), there are only the alternative manners — the sequence $\{C_1, \ldots, C_n\}$ is written as

$\{nC_1, (n-1)C_1, \ldots, C_1, (n-1)C_2, (n-2)C_2, \ldots, C_2, \ldots\ldots, 2C_{n-1}, C_{n-1}, C_n\}$,

or as

$\{C_{1,1}, C_{1,2}, \ldots, C_{1,n}, C_{2,1}, C_{2,2}, \ldots, C_{2,n-1}, \ldots\ldots, C_{n-1,1}, C_{n-1,2}, C_{n,1}\}$,

where

$$C_{1,1} = C_{1,2} = \ldots = C_{1,n} = C_1,$$
$$C_{2,1} = C_{2,2} = \ldots = C_{2,n-1} = C_2,$$
$$\ldots\ldots,$$
$$C_{n-1,1} = C_{n-1,2} = C_{n-1},$$
$$C_{n,1} = C_n.$$

It should be noted that the solution vector through the second manner might be possibly unstable.

Still take the sequence {211, 122, 300} and the equation $211x + 122y + 300z = 1177$.

By the first manner, they may be written respectively as {633, 422, 211, 244, 122, 300}, and as $633b_1 + 422b_2 + 211b_3 + 244b_4 + 122b_5 + 300b_6 = 1177$ with $b_i \in [0, 1]$. Then, the right solution vector to the rewritten equation is (1, 0, 0, 1, 0, 1) if it can be found out in polynomial time.

No matter what manner is adopted, the length of the extended sequence is $n + \ldots + 1 = n(n+1)/2$, and the size of the matrix corresponding to a lattice basis is $1 + n(n+1)/2$ by $1 + n(n+1)/2$. Hence, the density of the sequence $\{C_1, \ldots, C_n\}$ is in substance

$$\begin{aligned} D &= (n + \ldots + 1) / \log \max_{1 \le i \le n}\{C_i\} \\ &= n(n+1) / (2 \log \max_{1 \le i \le n}\{C_i\}) \\ &\approx n(n+1) / (2 \log M). \end{aligned}$$

For REESSE2+, $n \ge 120$ and $\log M \le 2n$, so $D \ge n(n+1)/(4n) = (n+1)/4 > 30 > 2$.

In light of the argumentation in section V.B.1, the ciphertexts in the REESSE2+ scheme with $D > 30$ and a large $n$ are secure against the lattice basis reduction attack.

*3) Cost of Reducing Ciphertexts via a $L^3$ Lattice Basis Is not Negligible:*

The set of all integral linear combinations of $n$ linearly independent vectors $\vec{b}_1, \ldots, \vec{b}_n \in \mathbb{R}^d$, as you see above, is called a lattice of dimension $n$. In the subset sum problem, it is easily understood that $d = O(n)$ frequently.

The original $L^3$ algorithm performs $O(n^5)$ arithmetic steps on $O(n^2)$-bit integers [9]. Owing to Gram-Schmidt orthogonalization, reduction process produces rational coefficients each of which is expressed by a $O(n^2)$-bit numerator and a $O(n^2)$-bit denominator. Hence, when $n$ is comparatively large, the $L^3$ algorithm is expensive for practical cryptanalytic applications.

If the rationals are cursorily substituted with floating-point numbers, the algorithm might not normally terminate, or the output basis might not be reduced. There was only provable floating-point $L^3$ algorithm known with a precision of $O(2n)$ bits [23]. In practice, people use its heuristic version [24]. By the cascade mode [25], the floating-point $L^3$ runs in $O(2n^7)$ bit operations.

In 2006, Schnorr brought forward Segment $L^3$-reduction under floating-point arithmetic [26]. $SLLL_0$ runs in $O(n^4)$ arithmetic steps with $O(n^2)$-bit floating-point numbers. SLLL in $O(n^4 \log n)$ arithmetic steps with $O(n)$-bit floating-point numbers. $SLLL^+$ in $O(n^3 \log n)$ arithmetic steps with $O(n^2)$-bit floating- point numbers. Obviously, the bit operations of SLLL, $O(n^6 \log n)$, is the lowest, but still laborious.

If the SLLL is employed for attacking the ciphertexts in the REESSE2+ system with $n \geq 120$, first, must consider acquiring the non-modular subset sum $\sum_{i=1}^{n} C_i b_i L_i$ from $\bar{E}$, which needs $O(n^2)$ heuristic trials, and second, must lay $n^2$ vectors in constructing a basis according to section V.B.2. On the basis of the two points, the bit operations of SLLL for REESSE2+ can be estimated at least at

$$\begin{aligned} & O(n^2)\, O((n^2)^6 \log n^2) \\ &= O(n^2 (n^2)^6 \log n^2) \\ &\approx O((2^7)^2\, (2^{7\times 2})^6 \log 2^{7\times 2}) \\ &\approx O(2^{14}\, 2^{84}\, 2^4) \\ &= O(2^{102}), \end{aligned}$$

and clearly, it is not negligible.

### C. Preventing Meet-in-the-middle Attacks

Let $b_1 \ldots b_n$ be a plaintext, $t = \lfloor n / 2 \rfloor$, and $S = \bar{E} + kM$ be a related ciphertext, where $k \in [0, n(n+1)/2]$.

Construct a table with entries

$$(\textstyle\sum_{i=t+1}^{n} C_i b_i L_i,\ (b_{t+1} \ldots b_n)),$$

try each combination of $b_1 \ldots b_t$, and judge whether

$$F = S - \textstyle\sum_{i=1}^{t} C_i b_i L_i$$

is the first component of some entry in the table. Therefore, the meet-in-the-middle attack method is likewise applicable for the REESSE2+ scheme [22].

Because an adversary needs $O(n^2)$ heuristic trials for acquiring the non-modular subset sum $S$ from $\bar{E}$, the cost of the meet-in-the-middle attack is $O(n^2) O(n 2^{n/2}) = O(n^3 2^{n/2})$ steps. Thus, when $n = 120$, this cost is approximately $O(2^{80})$ steps, and is enough for the security of a cryptosystem at present.

Also, it indicates that when $n = 120$, the number of bits which can be protected effectually by the REESSE2+ scheme is 80, and if the number of bits encrypted is greater than 80, an adversary should substitute the meet-in-the-middle attack for the brute force attack.

It is interesting that those bits after the bit string protected effectually exactly play a role resisting the adaptive-chosen -ciphertext attack.

### D. Avoiding the Adaptive-chosen-ciphertext Attack

Theoretically, absolute most of public key cryptographies may probably be faced by the adaptive-chosen-ciphertext attack. In 1998, Bleichenbacher demonstrated a practical adaptive-chosen-ciphertext attack on a form of RSA encryption [27].

In the same year, the Cramer-Shoup asymmetric encryption algorithm from the extremely malleable ElGamal algorithm was proposed [28]. It is the first efficient scheme proven to be secure against adaptive-chosen-ciphertext attack using standard cryptographic assumptions, which indicates that not all uses of cryptographic hash functions require random oracles — some require only the property of collision resistance.

An effectual approach to avoiding the adaptive-chosen-ciphertext attack is to append a stochastic fixed-length binary sequence to the terminal of every plaintext bock when it is encrypted. Thus, when a plaintext block is encrypted at different time with the same public key, the generated ciphertext blocks are distinct from one another. For example, a concrete implementation is referred to the OAEP+ scheme [29].

## VI. Analysis of Time Complexity of the REESSE2+ Scheme

Since the key generation algorithm is not required to be real-time, it is not intended to analyze the time complexity of this algorithm.

Hereunder, the time complexity of an algorithm is measured by the amount of bit operations (abo, shortly). Usually, the abo of a comparison operation is neglected. In terms of [22], the abo of a modular addition is $O(2 \log M)$, and the abo of a modular multiplication is $O(2 \log^2 M) = O(2(\log M)^2)$, where $M$ is a modulus.

### A. Time Complexity of the Encryption Algorithm

It is known from section III.B that the encryption algorithm has one loop, the loop body contains only a statement: $(\bar{E} + L C_i) \bmod M$, and the number of loop iterations is $n$.

Because $L \in [1, n]$ is extremely small, the bit operations of the multiplication in the expression $(\bar{E} + L C_i) \bmod M$ may be neglected, and the $(\bar{E} + L C_i) \bmod M$ may be regarded as modular addition arithmetic. In this way, the abo of encryption is $O(2n \log M)$, a linear function of $n$. Obviously, the encryption speed is extraordinarily fast.

### B. Time Complexity of the Decryption Algorithm

It is known from section III.C that the decryption algorithm contains two loops layered predominantly.

Step 1 contains a modular multiplication, and its abo is approximately

$$Mult = O(2 \log^2 M).$$

Step 2, …, and step 6 compose the outer loop body. It contains a modular addition and an inner loop. The abo of the modular addition is

$$Addi = O(2 \log M).$$

Let $\bar{U}$ denote the number of the outer loop iterations. The value of $\bar{U}$ rests with $L = L_1$, namely the number of 1-bits in a plaintext block and $\ell(1), \ldots, \ell(n)$, namely a distribution of integers $5, \ldots, n+4$. Note that this distribution is uniform. Due to the indeterminacy of $L_i$ and $\ell(i)$, we only can compute roughly the expected value of $\bar{U}$.

For convenience, we may substitute $\ell(i) \in [5, n+4]$ with $\ell(i)$

$\in [1, n]$.

Firstly, let $L = n$. Then

$$U_{\min} = (n)1 + (n-1)2 + \ldots 1(n) \\ = n(n+1)(n+2) / 6,$$

and

$$U_{\max} = n^2 + (n-1)^2 + \ldots + 1^2 \\ = n(n+1)(2n+1) / 6.$$

Let

$$k = U_{\max} - U_{\min} + 1 \\ = n(n+1)(n-1)/6 + 1 \\ = n(n^2-1)/6 + 1,$$

and $\bar{U}_n$ be the expected value as $L = n$. We have

$$\bar{U}_n = (t_1 U_{\min} + t_2(U_{\min}+1) + \ldots + t_{k-1}(U_{\max}-1) + t_k U_{\max}) / n!,$$

where $t_1 = t_k = 1$, the integers $t_2, \ldots, t_{k-1} \geq 0$, and $t_1 + t_2 + \ldots + t_k = n!$.

Since it is infeasible to compute $u_2, \ldots,$ and $u_{k-1}$, let

$$\bar{U}_n \approx (U_{\min} + U_{\max}) / 2 \\ = n(n+1)^2 / 4.$$

For instance, let $n = 3$ and $L = 3$. Then $L_1 = 3$, $L_2 = 2$, $L_1 = 1$, and $\ell(i) \in \{1, 2, 3\}$.

The enumeration of $(L_1, L_2, L_3) \times (\ell(1), \ell(2), \ell(3)) = L_1 \ell(1) + L_2 \ell(2) + L_3 \ell(3)$ is as follows:

$U_{\min} = (3, 2, 1) \times (1, 2, 3) = 10$, $(3, 2, 1) \times (1, 3, 2) = 11$, $(3, 2, 1) \times (2, 1, 3) = 11$, $(3, 2, 1) \times (2, 3, 1) = 13$, $(3, 2, 1) \times (3, 1, 2) = 13$, and $U_{\max} = (3, 2, 1) \times (3, 2, 1) = 14$.

Therefore, $\bar{U}_3 \approx (U_{\min} + U_{\max}) / 2 = (10 + 14) / 2 = 12$ while $\bar{U}_3 = (10 + 2 \times 11 + 2 \times 13 + 14) / (3 \times 2 \times 1) = 12$. Both the values are equal.

Likewise, when $n = 4$ and $L = 4$, $\bar{U}_4 \approx (U_{\min} + U_{\max}) / 2 = (20 + 30) / 2 = 25$ while $\bar{U}_4 = (20 + 3 \times 21 + 22 + 4 \times 23 + 2 \times 24 + 2 \times 25 + 2 \times 26 + 4 \times 27 + 28 + 3 \times 29 + 30) / (4 \times 3 \times 2 \times 1) = 25$. Both the values are also equal.

Secondly, let $L = n - 1$. Then

$$U_{\min} = (n-1)1 + (n-2)2 + \ldots + 1(n-1) \\ = n(n-1)(n+1) / 6.$$

For computational convenience, let

$$U_{\max} \approx (n-1)^2 + (n-2)^2 + \ldots + 1^2 \\ = n(n-1)(2n-1) / 6.$$

Therefore,

$$\bar{U}_{n-1} \approx (U_{\min} + U_{\max}) / 2 \\ = (n-1)\, n^2 / 4.$$

Similarly, we may obtain

$$\bar{U}_{n-2} \approx (n-2)(n-1)^2 / 4, \ldots, \text{and } \bar{U}_1 \approx 1(1+1)^2 / 4.$$

If the symmetric key $b_1 \ldots b_n \neq 0$ and $L \in [1, n]$ is uniformly distributed, the expected number of the outer loop iterations is

$$\bar{U} \approx (\bar{U}_n + \bar{U}_{n-1} + \ldots + \bar{U}_1) / n \\ = (n+1)(3n^2 + 11n + 10) / 48.$$

However, for a reasonable symmetric key, $L$ is equal to $n/2$ around. Therefore, the further reasonable expected value should be

$$\bar{U} \approx ((n+1)(3n^2 + 11n + 10) / 48 + \bar{U}_{n/2}) / 2 \\ = ((n+1)(3n^2 + 11n + 10) / 48 + n(n+2)^2 / 32) / 2 \\ \approx (n+1)(n^2 + 3n + 2) / 22.$$

When $96 \leq n \leq 176$, there is $\bar{U} \approx 6(n^2 + 3n + 2) \approx 6(n^2 + 3n)$.

At step 5, the inequality $E \leq E_i + L\dot{E}_i$ is a very strong constraint which causes that the expected number of the inner loop iterations is at most $n / 2$. Similar to section VI.A, the bit operations of the multiplication in the expression $E - LA_i$ may be neglected, and the $E - LA_i$ may be regarded as ordinary subtraction arithmetic. In this way, the abo of the inner loop is $Inlp = O(½\, n \log M)$.

By the algorithm, only if $\bar{E}$ is even and $\bar{E} \leq E_n$, is the inner loop executed. Obviously, the probability of $\bar{E}$ being an even number is 1 / 2. Additionally, in practice, there are $A_{i+1} / A_i \leq 3$ and $M / \sum_{i=1}^{n}(n+1-i)A_i \geq 3$, which means that probability of $\bar{E} \leq E_n$ is at most 1 / 3. Further, the probability that the two conditions are satisfied simultaneously by $\bar{E}$ is at most 1 / 6. Therefore, the abo of the decryption algorithm is

$$O(Mult + \bar{U}\, Addi + (1/6)\bar{U}\, InLp) \\ = O(2\log^2 M + 6(n^2 + 3n)(2\log M) + (n^2 + 3n)(½\, n \log M)) \\ \approx O(0.5n^3 \log M + 13.5n^2 \log M + 36n \log M + 2\log^2 M).$$

Clearly, it is a cubic function of $n$. Further, in the REESSE2+ system, due to $1.585n \leq \log M \leq 2n$, the abo of the decryption is at least $O(0.79n^4 + 21.39n^3 + 62n^2)$, and at most $O(n^4 + 27n^3 + 80n^2)$.

### *C. Comparison of REESSE2+ with ECC and NTRU*

To the REESSE2+ scheme, assume that $\log M = 1.6n$ in this section. Then, possible maximal $A_1$ is $3^m$, where $m$ satisfies $1.585(n + m) \approx 1.6n$. When $n = 120$, $m \approx 1$, and when $n = 176$, $m \approx 1.6$. Further, it is well understood that the length of a private key (lPvtk, shortly) is roughly $1.585n(2m + n - 1) / 2$ in bits, the length of a public key (lPubk, shortly) is $1.6n^2$ in bits, and the length of a ciphertext (lCiph, shortly) is $1.6n$ in bits. It follows that the abo of decryption is

$$O(0.8n^4 + 21.6n^3 + 63n^2).$$

It is known from section V.C that as $n = 120$ or $n = 176$, the security of REESSE2+ is equivalent to $2^{80}$ or $2^{112}$ steps, namely $2^{36}$ or $2^{68}$ mips years respectively.

Assume that $P \neq 2, 3$ is a prime, and $y^2 = x^3 + ax + b$ with $a, b \in \mathbb{GF}(P)$ is an elliptic curve. Then, according to [15] and [30], for the elliptic curve cryptosystem, the abo of encryption is roughly

$$O(40\log^3 P + 50\log^2 P + 10\log P),$$

and the abo of decryption roughly

TABLE I

COMPARISON OF REESSE2+ WITH ECC AND NTRU IN LENGTH

| | security (mips years) | modulus (bits) | *lPvtk* (bits) | *lPubk* (bits) |
|---|---|---|---|---|
| REESSE2+ /120 | $2^{36}$ | 192 | 11508 | 23040 |
| ECC /160 | $2^{36}$ | 160 | 160 | 640 |
| NTRU /251 | $2^{36}$ | 8 | 1004 | 2008 |
| REESSE2+ /176 | $2^{68}$ | 282 | 24856 | 49562 |
| ECC /224 | $2^{68}$ | 224 | 224 | 896 |
| NTRU /347 | $2^{68}$ | 9 | 1388 | 3123 |

TABLE II
COMPARISON OF REESSE2+ WITH ECC AND NTRU IN ABO

| | security (mips years) | *lCiph* (bits) | abo of encrypt. | abo of decrypt. |
|---|---|---|---|---|
| REESSE2+ /120 | $2^{36}$ | 192 | 51,840 | 204,120,000 |
| ECC /160 | $2^{36}$ | 640 | 165,121,600 | 82,947,200 |
| NTRU /251 | $2^{36}$ | 2008 | 65,789,108 | 131,578,216 |
| REESSE2+ /176 | $2^{68}$ | 282 | 111,514 | 887,319,910 |
| ECC /224 | $2^{68}$ | 896 | 452,088,000 | 226,800,000 |
| NTRU /347 | $2^{68}$ | 3123 | 172,574,204 | 345,148,408 |

$$\mathrm{O}(20\log^3 P + 40\log^2 P + 20\log P),$$

where some subordinate operations are ignored.

It is known from [30] that as $\log P = 160$ or $\log P = 224$, the security of EEC is equivalent to $2^{80}$ or $2^{112}$ steps, namely $2^{36}$ or $2^{68}$ mips years respectively.

In light of [31], the security of NTRU with $N = 251$, $q = 197$ and $d = 48$ is equivalent to that of ECC with $\log P = 160$, and NTRU with $N = 347$, $q = 269$ and $d = 66$ is equivalent to ECC with $\log P = 224$. In these two cases, the length of a plaintext block encrypted by NTRU is 80 bits and 112 bits respectively.

According to [32], the effort of the NTRU encryption is $4N^2$ additions and $N$ divisions by $q$. Generally speaking, the addition produces binary carry. Additionally, the effort of the decryption is roughly double that of the encryption. Hence, we see that the abo of encryption is roughly

$$\mathrm{O}(4N^2(N + \log q) + N \log q(N + \log q)),$$

and the abo of decryption is roughly

$$\mathrm{O}(8N^2(N + \log q) + 2N \log q(N + \log q)).$$

When the securities of REESSE2+, ECC and NTRU match reciprocally, their lengths and performances can be compared. Please see Table I and II.

The some facts can be observed from Table I and II. The size of a private key or a public key in REESSE2+ is longest, but is still tolerable. The length of a ciphertext in REESSE2+ is shortest, ECC secondary, and NTRU longest. The encryption speed of REESSE2+ is fastest, NTRU secondary, and ECC slowest. The decryption effort of REESSE2+ is roughly equivalent to NTRU and ECC since the ratios among them three all are not large from the angle of bit operation.

Furthermore, to satisfy extraordinary requirements under some circumstances, the bit-length of a private key or a public key can be shortened through a compression algorithm.

## VII. CONCLUSION

Resorting to the key transform $C_i \equiv (A_i + Z\ell(i))W$ (% $M$), REESSE2+ avoids the Shamir extremum attack. Resorting to an anomalous subset sum that is an extension of the connotation of the lever function, REESSE2+ avoids the $L^3$ basis reduction attack. Moreover, to place the multiple coefficient $L_i$ before every element in a set makes the subset sum have a direction, which can resists parallel arithmetic.

The mathematical implications of an anomalous super-increasing sequence and an anomalous sunset sum are not unique. For instance, we may define an anomalous super-increasing sequence with

$$A_i > \sum_{j=1}^{i-1} (i - j)^2 A_j,$$

and correspondingly an anomalous subset sum with

$$\bar{E} \equiv \sum_{i=1}^{n} C_i b_i L_i^2 \ (\%\ M),$$

where $L_i = \sum_{j=i}^{n} b_j$. In this case, the density of a public-key sequence will increase while the speed of the decryption algorithm will decrease.

The REESSE2+ public-key scheme is another application of the lever function and its connotation, holds a comparatively small modulus, and provides embedded computation platforms or mobile computation platforms with fast encryption and decryption implementations.

Similar to the REESSE1+ cryptosystem, to design a signature scheme on the basis of the REESSE2+ encryption scheme should be feasible.

## ACKNOWLEDGMENT

The authors would like to thank the Academicians Jiren Cai, Zhongyi Zhou, Jianhua Zheng, Changxiang Shen, Zhengyao Wei, Andrew C. Yao, Binxing Fang, Guangnan Ni, and Xicheng Lu for their important guidance, advice, and suggestions.

The authors also would like to thank the Professors Dingyi Pei, Jie Wang, Ronald L. Rivest, Moti Yung, Dingzhu Du, Mulan Liu, Huanguo Zhang, Dengguo Feng, Yixian Yang, Maozhi Xu, Qibin Zhai, Hanliang Xu, Xuejia Lai, Yongfei Han, Yupu Hu, Rongquan Feng, Ping Luo, Dongdai Lin, Jianfeng Ma, Lusheng Chen, Wenbao Han, Lequan Min, Bogang Lin, Hong Zhu, Renji Tao, Zhiying Wang, Quanyuan Wu, and Zhichang Qi for their important counsel, suggestions, and corrections.